# Nanoscale distribution of magnetic anisotropies in bimagnetic soft core-hard shell MnFe$_2$O$_4$@CoFe$_2$O$_4$ nanoparticles

*Niéli Daffé, Marcin Sikora, Mauro Rovezzi, Nadejda Bouldi, Véronica Gavrilov, Sophie Neveu, Fadi Choueikani, Philippe Ohresser, Vincent Dupuis, Dario Taverna, Alexandre Gloter, Marie-Anne Arrio, Philippe Sainctavit, and Amélie Juhin[*]*

N. Daffé
Institut de Minéralogie, de Physique des Matériaux et de Cosmochimie (IMPMC), UMR7590, CNRS, Université Pierre et Marie Curie, Institut de Recherche pour le Développement, 4 Place Jussieu, 75052 Paris Cedex 05, France.
Sorbonne Universités UPMC Univ Paris 06, UMR 8234, PHENIX F-75005 Paris, France.
Synchrotron SOLEIL, L'Orme des Merisiers, Saint-Aubin – BP48, 91192 Gif-sur-Yvette, France.
Dr. M. Sikora
AGH University of Science and Technology, Academic center for Materials and Nanotechnology, Al. Mickiewicza 30, 30-059 Krakow, Poland.
N. Bouldi, Dr. Ph. Sainctavit
Institut de Minéralogie, de Physique des Matériaux et de Cosmochimie (IMPMC), UMR7590, CNRS, Université Pierre et Marie Curie, Institut de Recherche pour le Développement, 4 Place Jussieu, 75052 Paris Cedex 05, France.
Synchrotron SOLEIL, L'Orme des Merisiers, Saint-Aubin – BP48, 91192 Gif-sur-Yvette, France.
Dr. M. Rovezzi
European Synchrotron Radiation Facility (ESRF), 6 Rue Jules Horowitz, BP220, 38043 Grenoble Cedex 9, France
Dr. V. Gavrilov, Dr. S. Neveu, Dr. V. Dupuis
Sorbonne Universités UPMC Univ Paris 06, UMR 8234, PHENIX F-75005 Paris, France.
Dr. F. Choueikani, Dr. Ph. Ohresser
Synchrotron SOLEIL, L'Orme des Merisiers, Saint-Aubin – BP48, 91192 Gif-sur-Yvette, France.
Dr. A. Gloter
Laboratoire de Physique des Solides, CNRS UMR 8502, Université Paris-Sud 11, 91405 Orsay, France.
Dr D. Taverna, Dr M.-A.Arrio, Dr. A. Juhin
Institut de Minéralogie, de Physique des Matériaux et de Cosmochimie (IMPMC), UMR7590, CNRS, Université Pierre et Marie Curie, Institut de Recherche pour le Développement, 4 Place Jussieu, 75052 Paris Cedex 05, France.
E-mail : amelie.juhin@impmc.upmc.fr

The nanoscale distribution of magnetic anisotropies was measured in core@shell MnFe$_2$O$_4$@CoFe$_2$O$_4$ 7.0 nm particles using a combination of element selective magnetic spectroscopies with different probing depths. As this picture is not accessible by any other technique, emergent magnetic properties were revealed. The coercive field is not constant in a whole nanospinel. The very thin (0.5 nm) $CoFe_2O_4$ hard shell imposes a strong magnetic anisotropy to the otherwise very soft $MnFe_2O_4$ core: a large gradient in coercivity was measured inside the $MnFe_2O_4$ core with lower values close to the interface region, while the inner core presents a substantial coercive field (0.54 T) and a very high remnant magnetization (90% of the magnetization at saturation).

1. Introduction

Ferrite nanoparticles with a spinel structure ($\gamma$-$Fe_2O_3$, $Fe_3O_4$, $MnFe_2O_4$, $CoFe_2O_4$) have become very popular for their original magnetic properties that are not shown by bulk materials. The reduced size combined to enhanced magnetic properties has led to numerous applications from biomedicine to high density storage devices,[1],[2],[3] although further developments and miniaturization of magnetic storage devices based on nanoferrites are now limited by the "superparamagnetic limit".[4] To overcome this, chemists have imagined new synthesis pathways to tune further the magnetic properties of nanospinels by playing on their structuration, leading to the emergence of core@shell nanostructures which rapidly focused interest.[5] Since they allow the possibility to combine a core and a shell from two different magnetic species (such as for example a ferro/ferrimagnetic phase with an antiferromagnetic one,[6],[7],[8] or a soft magnetic with a hard magnetic one[9],[10],[11]), bimagnetic core@shell nanospinels offer a fertile ground to overtake the limits of conventional magnetization parameters (anisotropy constant $K$, blocking temperature $T_B$, saturation magnetization $M_s$ and coercive field $H_c$ [12],[13],[14]) or to observe strong exchange coupling between both magnetic phases, such as exchange bias[15],[16],[17] or exchange-spring[18],[19] coupling effects. In parallel, magnetic nanoparticles have found numerous applications when dispersed in liquids to form stable magnetic colloids.[20] Such systems are called ferrofluids when they are composed of ferro- or ferrimagnetic single domain nanoparticles, with a typical diameter in the 5 nm-30 nm range.[21] In biomedicine, ferrofluids of core@shell nanoparticles have recently been considered for advanced applications, such as nanoheater-based therapies for cancer treatment (magnetic hyperthermia), which use radiofrequency fields.[22],[23],[24] The efficiency of the process relies on the magnetic anisotropy of the nanoparticle, which governs the high frequency losses, hence the idea to modulate magnetic anisotropy by designing novel core@shell architectures and inducing emergent magnetic properties.

In this work, we study the magnetic anisotropies in well-crystallized 7.0 nm core@shell nanoparticles, which combine a soft core of $MnFe_2O_4$ covered with a thin (0.5 nm) hard shell of $CoFe_2O_4$. Using an original experimental approach combining electron and x-ray techniques with different probing depths, we investigate separately the cationic distribution and magnetic anisotropies in the core and those in the shell, as well as their mutual influence. Among the techniques dedicated to study magnetic nanoparticles, X-ray Absorption Spectroscopy (XAS) and X-ray Magnetic Circular Dichroism (XMCD) have now become standard techniques.[25],[26],[27],[28] XMCD is the difference in the absorption of circularly polarized X-rays for opposite sample magnetization direction. It can be measured in ferro- and ferrimagnetic materials and provides information on the magnetic properties of the absorbing atom.[29],[30]

XMCD measurements are mainly performed using soft X-ray photons (~600 and 780 eV at the Mn and Co $L_{2,3}$ absorption edges, respectively) delivered by synchrotron radiation facilities. Thanks to element specificity, XMCD allows disentangling the magnetic signature of the core - by measuring the Mn absorption edge, from that of the shell - by measuring the Co absorption edge. The existence of a possible magnetic coupling between both components can be determined.[31],[32] Additionally, the cationic distribution of the different ions amongst crystallographic sites can be solved thanks to the sensitivity of the XAS and XMCD spectral shape to the local site symmetry of the absorbing atom and to its valence, which yields crucial information to rationalize the magnetic properties of these materials. However, for core@shell

nanoparticles, where the magnetic structure of the outermost layers is usually very different from that of the inner core, a quantitative analysis based on XAS and XMCD is often limited by the short penetration depth of soft X-rays. Indeed, using the total electron yield traditional detection mode, the probing depth is rather small, with 60% of the signal coming from the top 2 nm. In order to circumvent this issue, hard X-rays with a much larger penetration depth can be used to probe the full nanoparticle depth. In addition to soft X-ray XAS and XMCD measurements, we have investigated $MnFe_2O_4@CoFe_2O_4$ particles using the RIXS-MCD technique, which consists in coupling hard X-ray MCD and Resonant Inelastic X-ray Scattering (RIXS) spectroscopy at the *K*-edge of *3d* ions.[33],[34] RIXS-MCD was recently used to investigate the buried interface in bi-magnetic core@shell $\gamma$-$Fe_2O_3$@$Mn_3O_4$ nanoparticles.[35] Such a combined approach is, to our knowledge, the first investigation of its kind.

## 2. Results and Discussion

### 2.1. Structural properties of dried nanoparticles

Two types of nanoparticles were synthesized as ferrofluids in heptane (see Experimental Section): $MnFe_2O_4$ and $MnFe_2O_4@CoFe_2O_4$ core@shell (also labeled Mn@Co hereafter). The latter were synthesized using the seeded-growth approach, *i.e.,* Mn@Co particles are built up from the pre-made nanoparticle core of $MnFe_2O_4$ sample on which a shell of $CoFe_2O_4$ is grown. Atomic Absorption Spectrometry measurements provide an estimate of the metallic molar concentration in each sample (**Table 1**), showing that the obtained stoichiometry is close to the expected one for a spinel, *i.e.* 1 divalent ion for 2 trivalent ions. The TEM micrographs illustrated in **Figure 1** show that all particles are spherical. The respective particle size histograms are well-fitted using normal distributions, leading to a mean diameter of 6.0 nm ($\sigma = 0.25$) for $MnFe_2O_4$ and 7.0 nm ($\sigma = 0.31$) for $MnFe_2O_4@CoFe_2O_4$,

respectively (Table 1). These values are consistent with the size of coherent domains obtained from the XRD patterns (**Figure S1**, Supporting Information), which also confirm the spinel structure of both samples. From the difference in particle size, the thickness of the outer cobalt ferrite layer in the MnFe$_2$O$_4$@CoFe$_2$O$_4$ particles can be estimated to be ~0.5 nm.

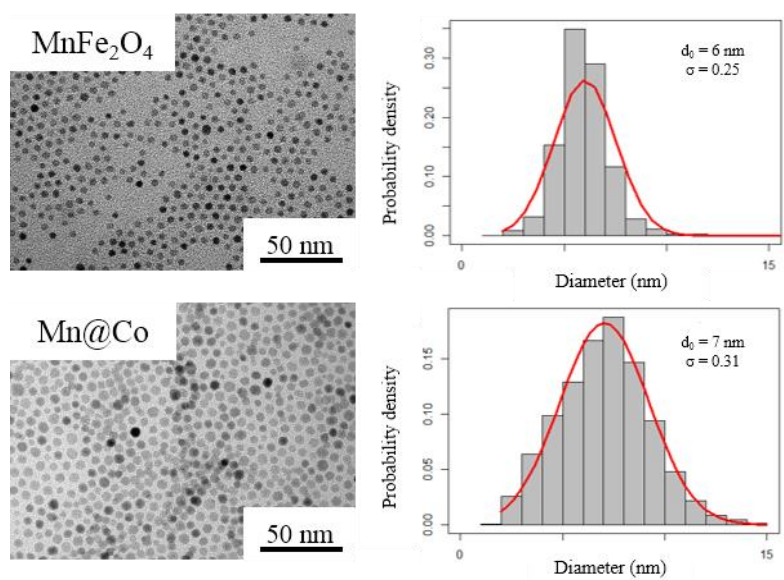

**Figure 1**. TEM images and distribution in particle size for the MnFe$_2$O$_4$ (top), and Mn@Co dried powder samples (bottom). Size distributions are determined using a Gaussian fit of histograms (red line).

High Resolution Scanning Transmission Electron Microscopy (STEM) and Electron Energy Loss Spectroscopy (EELS) analysis were performed to assess the elemental distribution within a Mn@Co nanoparticle, using a 0.1 nm step size and a reduced electron probe with 0.2 nm width. Results are shown in **Figure 2**. The High Angle Annular Dark Field (HAADF) signal acquired in parallel to the spectral information shows that the particle has a slightly faceted morphology (Figure 2a). The spatial distribution of Mn, Fe and Co ions obtained from the STEM-EELS analysis is mapped in Figures 2b-2e. The core@shell structuration of the nanoparticle is clearly visible, with Co ions exclusively located in the shell region and Mn ions mostly confined in the core region. This confirms the existence of the (Co, Fe)-rich outer shell grown on the (Mn, Fe)-rich core, but the local presence of residual Mn in the outer shell cannot be totally excluded. Indeed, when comparing the EELS spectrum for two selected areas that correspond to the inner core and the outermost shell region (labeled (A) and (B), respectively, Figure 2f), the presence of a residual signal at the energy loss of the Mn $L_{2,3}$ edges cannot be discarded for area B, but the signal of Co is clearly more intense. Intermixing between the $MnFe_2O_4$ core and the $CoFe_2O_4$ shell is therefore very likely but seems limited.

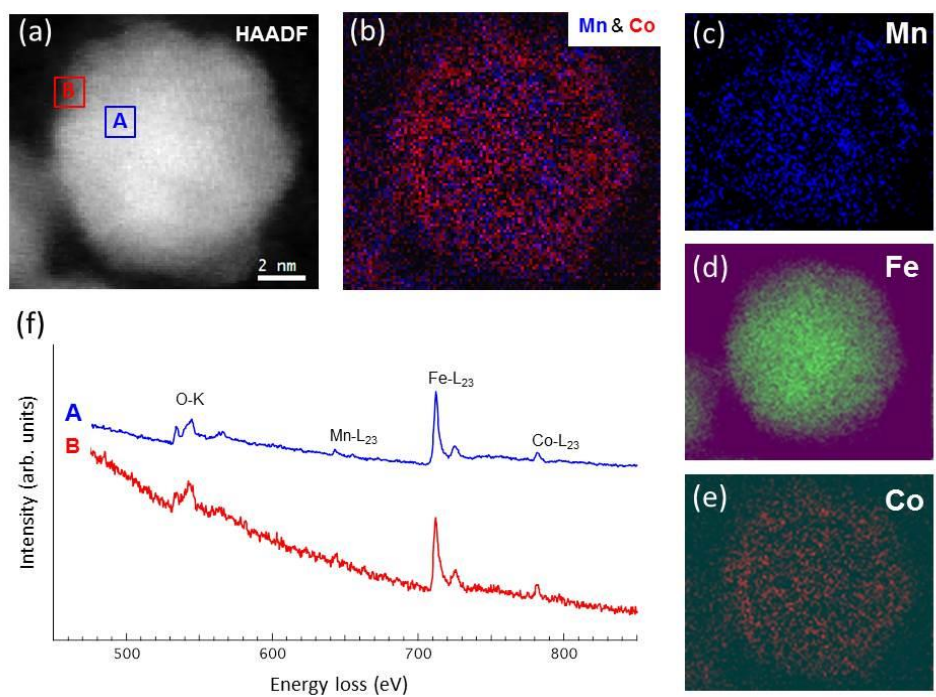

**Figure 2.** STEM-EELS analysis on an individual core@shell nanoparticle: (a) HAADF image; (b) superposition of the elemental maps of Mn and Co; (c) elemental map of Mn ; (d) elemental map of Fe; (e) elemental map of Co; (f) EELS spectra corresponding of selected areas in the core (spectrum A) and in the outer shell regions (spectrum B).

The temperature dependence of magnetization was measured using a Vibrating Sample Magnetometer (VSM) magnetometer. The value of the superparamagnetic blocking temperature ($T_B$) is estimated from the maximum of the Zero Field Cooled (ZFC) curve plotted in **Figure 3** together with the Field Cooled (FC) temperature dependence of magnetization.[36],[37] The Mn@Co nanoparticles show a higher blocking temperature ($T_B$ = 150K) than the $MnFe_2O_4$ nanoparticles ($T_B$ = 20K). The blocking temperature measured for MnFe$_2$O$_4$ nanoparticles is typical for such particle size and composition.[38] Since $T_B = \frac{KV}{25k_B}$ (where $K$ is the magnetic anisotropy, $V$ the volume and $k_B$ the Boltzmann constant), the increase of $T_B$ for the Mn@Co nanoparticles with respect to the MnFe$_2$O$_4$ seeds may be attributed to the slight volume increase due to the growth of the shell and most likely to the increase in magnetic anisotropy due to the formation of the harder CoFe$_2$O$_4$ outer layer. In order to disentangle both effects, we have investigated the depth-dependence of magnetic anisotropies inside a nanoparticle.

**Table 1.** Structural and magnetic properties of the MnFe$_2$O$_4$ and Mn@Co nanoparticles.

|  | MnFe$_2$O$_4$ | Mn@Co |
|---|---|---|
| Particle mean diameter from TEM (nm) | 6.0 | 7.0 |
| Particle mean diameter from XRD (nm) | 5.7 | 6.3 |
| Molar fraction of divalent metal $X_M$ (respectively, Mn, and (Co+Mn)) | 0.31 | 0.24 |
| Blocking temperature $T_B$ (K) from VSM | 20 | 150 |

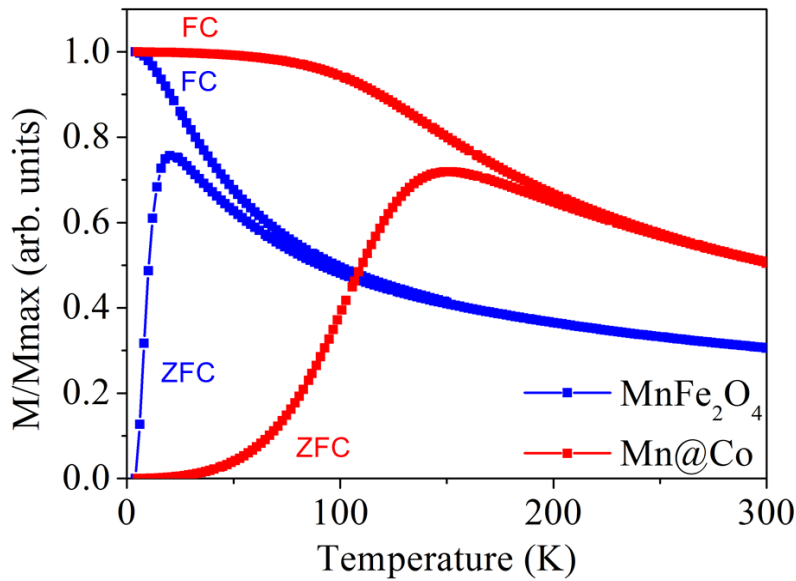

**Figure 3.** Field Cooled (with a probing field of 70 Oe) and Zero Field Cooled curves measured on MnFe$_2$O$_4$ (blue squares) and Mn@Co nanoparticles (red squares).
**2.2. Electronic structure and cationic repartition.**

Two types of measurements were performed at low temperature using x-ray magnetic spectroscopies. In a first step, spectral signatures related to the structural, electronic and magnetic properties were recorded. In a second step, element-selective magnetization *versus* field curves were measured in order to access the values of coercive field and remnant magnetization. Both types of measurements were conducted with surface sensitive XAS/XMCD (soft x-rays) and then with bulk sensitive RIXS/RIXS-MCD, therefore providing information with different probing depths. RIXS is a *photon-in, photon-out* element selective X-ray spectroscopy that provides for the absorbing atom a unique bi-dimensional mapping of its spectral signature.[39] The latter is intimately related to its electronic structure, *i.e.,* mainly its valence (number of *3d* electrons) and site symmetry (tetrahedral or octahedral).

In **Figure 4** are shown the RIXS and RIXS-MCD planes measured on the Mn@Co particles at low temperature (*i.e.,* below the particle blocking temperature and the solvent freezing point) using an external magnetic field of 1.2T, at the Mn *K*-edge (panels (a) and (b)) and at the Co *K*-edge (panels (c) and (d)). This allows probing selectively the speciation of Mn ions (that are located in the $MnFe_2O_4$ core) and that of Co (that are expected in the $CoFe_2O_4$ shell). At the Mn edge, the large intensity of the RIXS-MCD signal (~20% peak-to-peak) and its characteristic shape (one positive peak at lower incident energy and lower energy transfer, accompanied by one negative peak) reveal the dominant contribution from tetrahedral $Mn^{2+}$ ions, whose 2D spectral signature is also similar to that of tetrahedral $Fe^{3+}$ ions (which are isoelectronic to $Mn^{2+}$ ions) in $Fe_3O_4$ and $\gamma$-$Fe_2O_3$.[34] Using the same polarization and field conventions as in the latter experiments, the sign of the RIXS-MCD signal indicates that $Mn^{2+}$ ions have their magnetic moments opposite to the field direction, which further confirms their location in the tetrahedral sites of the $MnFe_2O_4$ direct spinel structure. At the Co edge, the small peak-to-peak intensity of the RIXS-MCD signal (~4%) and its shape, which is dominated by two negative features, indicate the presence of $Co^{2+}$ ions in the octahedral sites of the $CoFe_2O_4$ direct spinel structure.

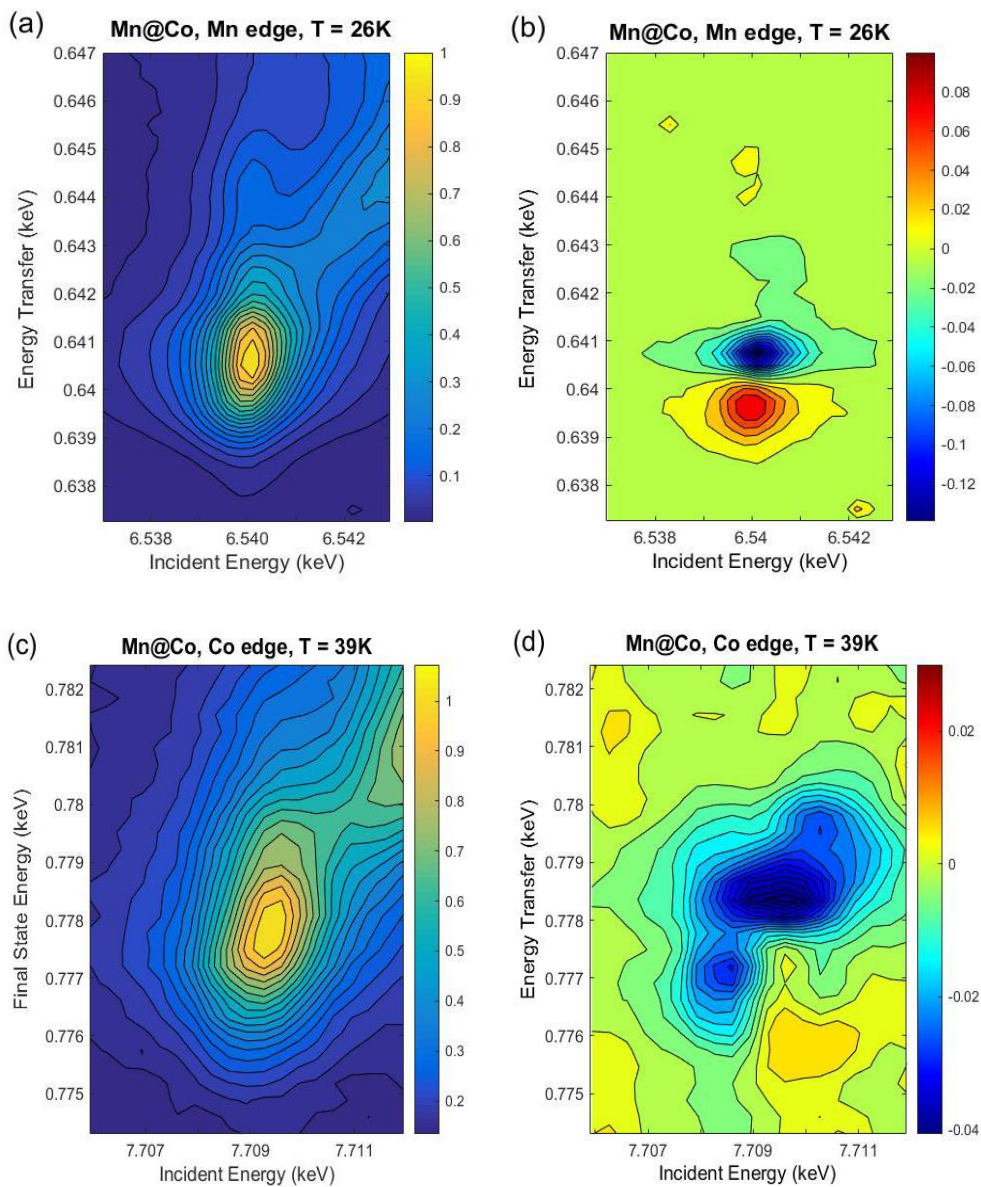

**Figure 4.** Low temperature RIXS and RIXS-MCD planes measured on Mn@Co particles at the Mn edge (panels (a) and (b)) and at the Co edge (panels (c) and (d)). For both edges only the part of the planes corresponding to the $K\alpha_1$ emission region is shown.

The soft XAS/XMCD measurements provide similar outcomes regarding the speciation of Mn and Co ions inside a Mn@Co particle, although the much smaller probing depth (2 nm vs a few μm) informs only on the upper part of the particle (*i.e.,* the Co shell and the Mn ions at the core-shell interface): the XMCD spectra measured at the different metal $L_{2,3}$ edges (**Figure S2**, Supporting Information) show that 100% of Co ions are divalent and in octahedral sites,[40] while Mn ions are found exclusively divalent and mainly in tetrahedral sites.[41]

**2.3. On the intermixing and the magnetic coupling between the core and the shell.**

The RIXS and RIXS-MCD signatures measured in the Mn@Co ferrofluid were compared with those measured in the parent $MnFe_2O_4$ particles at the Mn edge. In an attempt to compare the spectral signature measured at the Co edge with that of a known reference sample, 6 nm $CoFe_2O_4$ nanoparticles were prepared with a similar synthesis route and measured (See Experimental Section and Supplementary Information, **Figures S3, S4 and S5**). The comparison of spectral signatures can be made on the 2D planes (Figure 4, **Figures S6 and S7**, Supporting Information) which contain the full information, or more conveniently, on 1D spectra extracted from the 2D planes, either as Constant Emission Energy (CEE) scans (a diagonal cut in the plane, **Figure 5**) or as Constant Incident Energy (CIE) scans (a vertical cut in the plane, **Figures S8 and S9**, Supporting Information). The spectral signature of Mn ions in the Mn@Co particles, which arises from the $MnFe_2O_4$ core, is very similar to that of the Mn ions in the $MnFe_2O_4$ particles. Likewise, Co ions in the $CoFe_2O_4$ particles and those of the $CoFe_2O_4$ shell in the Mn@Co sample show MCD spectra that are identical within noise level (note that the signal from the 0.5 nm thin shell is very weak). This again confirms that in Mn@Co particles, Mn ions are mainly located in tetrahedral sites and Co ions in octahedral sites.

From the MCD spectra one can determine the magnetic coupling between the core and the shell. Since the magnetic moments of tetrahedral $Mn^{2+}$ ions are opposite to the external magnetic field direction and those of octahedral $Co^{2+}$ ions are along the field direction, we can conclude that the coupling between the core and the shell is ferromagnetic in nature. In other words, magnetic moments in the tetrahedral sites of the core and in those of the shell are all antiparallel to the external magnetic field, while those in the octahedral sites of the core and the shell are all parallel to the external field.

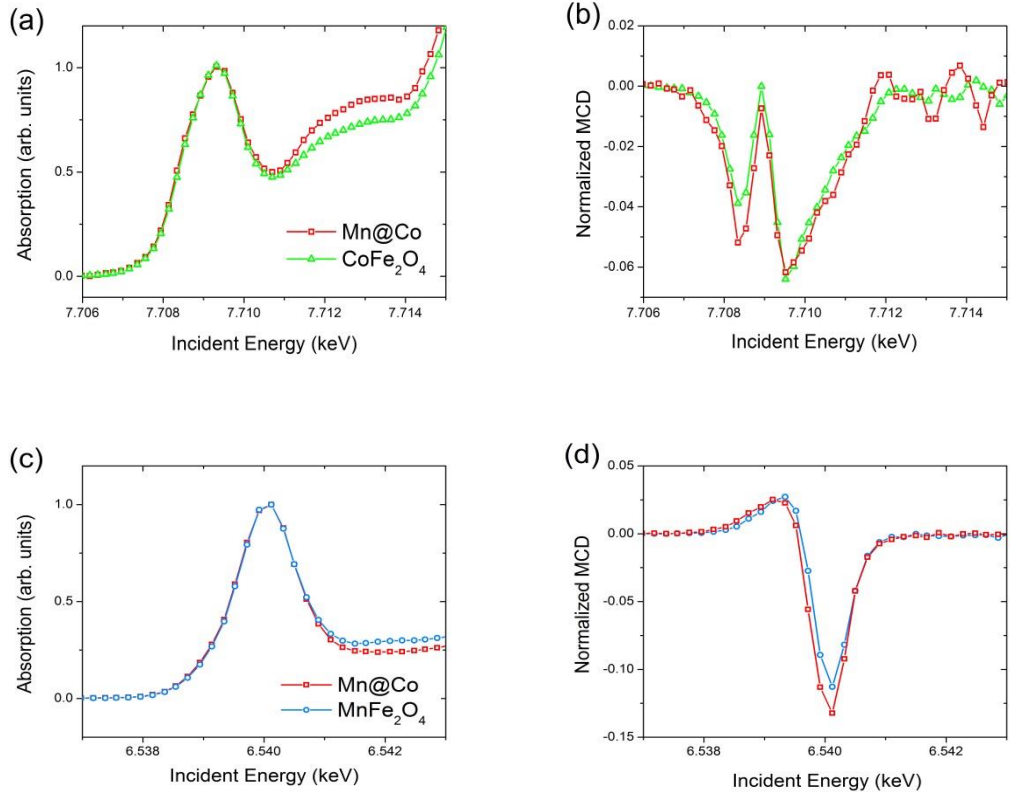

**Figure 5**. Low Temperature absorption and MCD spectra measured at Constant Emission Energy at the Co edge (panels (a) and (b), T=39K, CEE =6.9313 keV) and at the Mn edge (panels (c) and (d), T=26K, CEE=5.8998 keV). (Note that due to experimental issues it was not possible to measure at the very same temperature for both edges).

## 2.4. Element specific magnetic properties of the shell and the core.

In **Figure 6** are shown the magnetization *vs* magnetic field curves acquired at 26 K with a PPMS instrument for $MnFe_2O_4$ and Mn@Co particles. At this temperature, a very low coercive field of ~1mT is measured for $MnFe_2O_4$ nanoparticles while it is two orders of magnitude larger for Mn@Co nanoparticles (~0.5 T), which is in line with the soft nature of $MnFe_2O_4$. However, disentangling quantitatively the respective contributions of both magnetic components to the resultant average magnetization curve is tedious; hence our alternative approach to measure MCD detected magnetization curves that provide element selectivity.

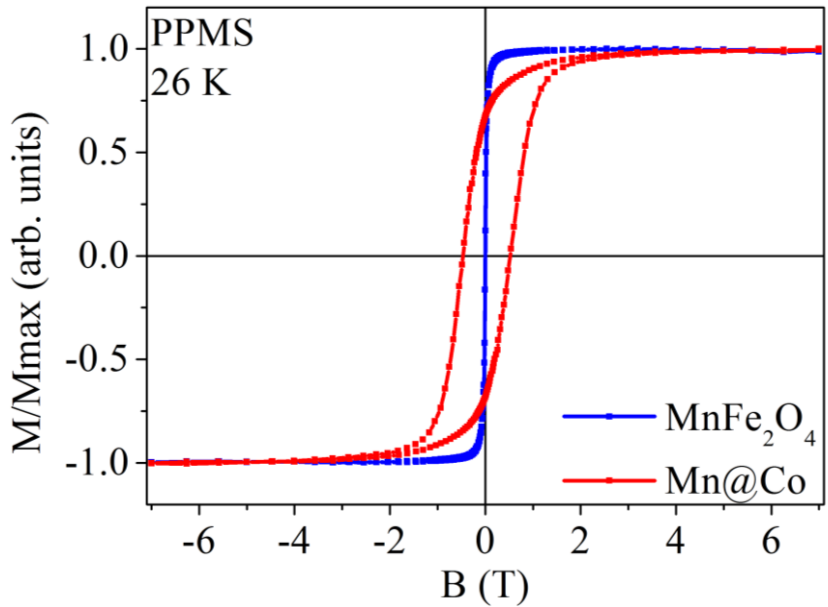

**Figure 6.** Magnetization *versus* field measurements with PPMS at T=26K for $MnFe_2O_4$ and Mn@Co particles.

Such element selective magnetization measurements were performed below the blocking temperature of Mn@Co nanoparticles, first at the Mn edge and then at the Co edge. MCD detected magnetization curves are measured by setting the incident energy (and also the emission energy for RIXS) to the value that maximizes the MCD intensity and by sweeping the magnetic field. X-rays with different penetration depths were used, either hard X-rays (RIXS-MCD) penetrating the whole nanoparticle depth, or soft X-rays (XMCD) with Total Electron Yield detection probing mainly the top 2 nm: this combination provides complementary information on magnetic anisotropies arising respectively from the whole particle and from its surface only. Results are shown in **Figure 7**. At T=39K, the coercive field measured at the Co edge with RIXS-MCD is $H_c$=0.18 Tesla and is consistent with the coercivity measured at the Co edge using soft XMCD ($H_c$=0.15 T, left panel). Although both values indicate the hard magnet behavior of the $CoFe_2O_4$ shell, they are significantly lower than the coercive field measured for the 6 nm $CoFe_2O_4$ reference particles ($H_c$=0.32 Tesla, **Figure S10**, Supporting Information). This difference is due to the fact that the latter have a very different magnetic anisotropy from that of a $CoFe_2O_4$ shell with 0.5 nm thickness, hence the necessity to measure the specific magnetic properties of the shell in the Mn@Co particles rather than using those of reference samples.

The normalized remnant magnetization (remnant magnetization divided by magnetization at saturation) is only ~50%, which could be the result of strong magnetic spin canting for the Co ions in the Mn@Co nanoparticles. Indeed, the canting is expected to be rather large because of the high proportion of Co ions situated at the particle surface that have less magnetic neighboring atoms than inside the bulk.[42] In a magnetic spinel structure, the magnetic structure is mainly governed by the antiferromagnetic coupling between tetrahedral and octahedral sites. Nevertheless, in a series of spinels bearing $Fe^{2+}$ and $Fe^{3+}$ in octahedral sites, it was shown that the magnetic interaction between octahedral sites is antiferromagnetic between $Fe^{3+}$ and $Fe^{3+}$ (J=-1.4K) and between $Fe^{2+}$ and $Fe^{2+}$ (J=-3.3K) because of super exchange, but it is ferromagnetic between $Fe^{2+}$ and $Fe^{3+}$ due to double exchange (J=+1.6K).[43] Both super exchange interactions plus the double exchange yield an average antiferromagnetic coupling between octahedral sites. At the surface of the Mn@Co particles, if the dominant antiferromagnetic coupling between tetrahedral and octahedral sites is reduced due to the finite size of the particle, the average antiferromagnetic coupling between octahedral sites now competes efficiently with the antiferromagnetic coupling between octahedral and tetrahedral sites. This results in the breaking of the collinear spin configuration.[44],[45]

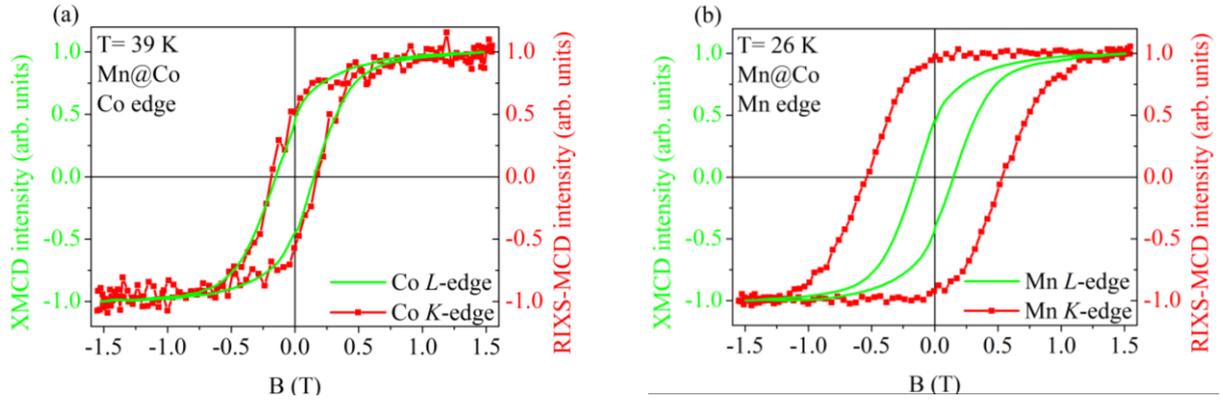

**Figure 7.** Element selective magnetization curves measured in Mn@Co at low temperature by XMCD using soft X-ray MCD (green lines) and hard X-ray RIXS-MCD (red lines), at the Co edge (left panel) and at the Mn edge (right panel).

At T=26K, the coercivity measured with hard X-rays ($H_c^{BS}$, where BS means Bulk Sensitive) at the Mn edge in the Mn@Co particles is 0.54 Tesla, while it is almost zero (0.013 Tesla) in the MnFe$_2$O$_4$ sample (**Figure S11**, Supporting Information). The normalized remnant magnetization is ~90 +/- 5%, which is similar to the expected value of 83-87% for randomly spherical ferrite nanoparticles with cubic anisotropy.[46] Such a high value associated to the square (non-slanted) shape of the magnetization curve reveals the lack of magnetic spin canting for the Mn ions. This is consistent with most of the Mn ions being in the "bulk" of the particle rather than partially diluted in the shell. In addition, since Mn ions are on the $T_d$ sites, they tend to be much less sensitive to spin canting that is preferentially affecting the $O_h$ sites.[45],[36]

Nevertheless, the coercivity induced in the core by the shell ($H_c^{BS}(Mn)$ = 0.54 Tesla at T= 26K) is much larger than the coercivity of the shell itself ($H_c^{BS}(Co)$ = 0.18 Tesla at T = 39K, a value that is likely similar at T = 26K because of the low T-dependence of $H_c$ measured at these temperatures by XMCD at the $L_{2,3}$ edges, see **Figures S12 and S13**, Supporting Information). This seems at first surprising, because one would expect the coercive field in the shell to induce a coercivity of the same magnitude in the single domain core. Instead, when the magnetic field is set to the value of $H_c^{BS}(Co)$, leading to zero magnetization on the shell, a significant remnant magnetization is still observed on the core. The difference between the coercivity measured in the shell and the one measured in the core is difficult to explain by the existence of several populations of magnetic Co ions. Indeed, attempts to reproduce the shape of the hysteresis loop measured at the Co edge in the Mn@Co sample using one population of Co ions that would follow the magnetization curve of the Mn ions, plus one population of paramagnetic Co ions (as given by a Brillouin curve for S=3/2), did not succeed. Instead, the difference between $H_c^{BS}(Co)$ and $H_c^{BS}(Mn)$ may arise from a gradient in coercivity inside the MnFe$_2$O$_4$ core. Indeed, using the surface sensitive XMCD probe, one finds that the coercivity ($H_c^{SS}$, SS means Surface Sensitive) of the probed Mn ions ($H_c^{SS}(Mn)$=0.15 Tesla at 26K and $H_c^{SS}(Mn)$=0.12 Tesla at 39K, Figure S9, Supporting Information), which are those located near the core-shell interface, is comparable to that of the Co ions in the shell ($H_c^{SS}(Co)$= 0.16 Tesla at 26K and $H_c^{SS}(Co)$=0.15 Tesla at 39K, see Figure S13, Supporting Information, which is also close to $H_c^{BS}(Co)$=0.18 Tesla at T=39K). Comparison with the coercive field measured for the entire particle at the Mn edge ($H_c^{BS}(Mn)$ =0.54 Tesla) implies the existence of a strong gradient in coercivity inside the MnFe$_2$O$_4$ core: lower values of coercive field arise from the interface region, while higher values originate from the center of the core.

### 3. Conclusion

Emergent magnetic properties were measured in core@shell MnFe$_2$O$_4$@CoFe$_2$O$_4$ 7.0 nm particles using a combination of element selective magnetic spectroscopies with different probing depths. A strong gradient in coercivity was measured inside the MnFe$_2$O$_4$ core with lower values close to the interface region (0.15 T at 26K), while the inner core presents a rather large coercive field (0.54 T at 26K). The coercivity of the inner core is larger than both the one of the thin 0.5 nm CoFe$_2$O$_4$ shell (~0.16 T) and the one of bare MnFe$_2$O$_4$ particles (0.01 T). In addition, our microscopic and spectroscopic findings show that only a limited interdiffusion is occurring, with no sign of cationic rearrangement that would explain such a spectacular change in anisotropy. This is clear evidence that the magnetic properties of the investigated core@shell particles arise from an emergent magnetic coupling between the core and the shell, which is due to the presence of the interface and is found to be ferromagnetic in nature. We expect that micromagnetic atomistic simulations could allow understanding the mechanism of the magnetic interaction between the core and the shell.

4. **Experimental Section**

*Synthesis.* Nanoparticles were prepared following the thermal decomposition process of metallic acetylacetonate precursors in the presence of oleic acid, 1,2-hexadecanediol, oleylamine and benzyl ether. Nanoparticles with controlled morphology and excellent crystallinity were obtained following the process of Sun *et al*.[47] The synthesis of $MnFe_2O_4$, $CoFe_2O_4$ and $MnFe_2O_4@CoFe_2O_4$ nanoparticles has been described elsewhere.[48]

*Transmission Electron Microscopy.* TEM micrographs were obtained using a JEOL 100 CX2. Size histograms were obtained from the analysis of the TEM micrographs over more than 10,000 nanoparticles. These histograms were fitted by a normal law with a least-square refinement that provides particles mean diameter $d_0$ and their respective polydispersity $\sigma$.

*X-Ray Diffraction.* XRD $\theta/2\theta$ patterns were recorded with a PANALYTICAL X'Pert Pro MPD diffractometer using Fe filtered the Co K$\alpha$1 ($\lambda$ = 1.79Å) radiation from a mobile anode at 40 kV, 40 mA. The measurements were carried out in a range of 30-80° $2\theta$ in steps of 0.02° and collection time of 7200s. Coherent domain sizes are calculated from the Scherrer equation.[49]

*Atomic Absorption Spectrometry.* Measurements were performed using a Perkin Elmer Analyst 100 with an air-acetylene flame at a mean temperature of 2300°C. Experiments were repeated at least three times on each sample. Molar concentrations of cobalt, manganese and iron are determined after degradation of the nanoparticles in concentrated hydrochloric acid. From the metallic concentrations species, the volume fraction of the nanoparticles was calculated and molar ratios $X_M$ are determined for each sample (Table 1): $X_M$ = [M] / ([M] + [Fe]) where [M] is [Mn] for $MnFe_2O_4$ and [Mn] + [Co] for Mn@Co.

*EELS.* High resolution Scanning Transmission Electron Microscopy and Electron Energy Loss Spectroscopy (EELS) analysis were performed using a Cs aberration-corrected STEM, the NION UltraSTEM200 operated at 100 kV and coupled with a high-sensitivity EEL spectrometer. The transmitted electron beam at each probed position has been analyzed with the spectrometer, to obtain a spectrum-image. The total dataset is formed by 109x96 pixels (spectra) with a beam spot size of 0.2 nm, and the acquisition time of each spectrum is 10 ms.

*Bulk magnetic measurements.* They were performed using a Vibrating Sample Magnetomer Quantum Design PPMS. Blocking Temperatures were estimated from the Zero-Field Cooled and Field Cooled (ZFC/FC) temperature dependence of magnetization measured under a 70 Oe field on the particles dispersed in paraffin. Paraffin is used here to avoid the critical fusion point of heptane. Magnetization curve vs. Field measurements were performed at 26K on the ferrofluids of the $MnFe_2O_4$, $CoFe_2O_4$ and $MnFe_2O_4@CoFe_2O_4$ particles.

*Soft XAS and XMCD measurements.* XAS and XMCD signals were recorded at the Mn and Co $L_{2,3}$ edges on the DEIMOS beamline at the French synchrotron, SOLEIL. The resolving power E / ΔE is better than 5000. Spectra were measured in Total Electron Yield at 50K and in High Vacuum conditions ($10^{-8}$ mbar) on a dropcast of Mn@Co particles. XMCD signals were recorded by flipping both the circular polarization (either left or right helicity) and the external magnetic field (either +1.5 Tesla or −1.5 Tesla). Circularly polarized X-rays are provided by an Apple-II HU52 helical undulator for both XAS and XMCD measurements. Element-specific magnetization curves were measured using the EMPHU-65 undulator which allows the fast switching (1 Hz) of the X-ray helicity.[50]

*RIXS and RIXS-MCD spectroscopies.* Experiments were carried out at ID26 beamline of the European Synchrotron Radiation Facility (Grenoble, France). Measurements were performed at the Co and Mn *K*-edges at low temperature on the frozen phase of ferrofluids using a dedicated liquid cell. The uncertainty on temperature is estimated to be +/- 5K. The incident energy was selected using a pair of Si(311) crystals. The intensity of Mn and Co *Kα* emission lines (of the inelastically scattered beam) was analyzed using a set of four spherically bent Ge(111) and Si (531) crystals, respectively, arranged with an Avalanche Photo Diode in the Rowland Geometry with a scattering angle of 90°. The overall resolution was measured at 0.7 eV and 0.8 eV for Mn and Co respectively. For both Co and Mn edges, *1s2p* RIXS planes were recorded as a set of Constant Energy Transfer scans over the energy of the $K\alpha_1$ line and of the *K* pre-edge. Additionally, absorption spectra were measured using High Energy Resolution Fluorescence Detection (HERFD) by setting the emission energy to the maximum of the *Kα* line (6.9313 keV for Co, 5.8998 keV for Mn), namely HERFD-XAS. HERFD-MCD and RIXS-MCD experiments were carried out with the same setup as for HERFD-XAS and RIXS measurements, the differences being that *(i)* the incident beam is circularly polarized (instead of linearly polarized in the case of HERFD/RIXS), *(ii)* samples are kept in magnetic saturation using an electromagnet allowing to reach a magnetic field of 1.5 T, for which a detailed calibration curve was measured. The circular polarization was obtained using a 500 μm thick diamond (111) quarter wave plate set downstream the Si(311) monochromator, with a circular polarization degree estimated to be 75 %. RIXS-MCD planes were recorded by reversing the photon helicity at each incidence energy. RIXS-MCD and HERFD-MCD spectra were recorded in the region of the Co and Mn *K* pre-edge, after freezing the samples with no external magnetic field Spectra were normalized such that the maximum of the polarization-averaged absorption spectrum is equal to 1 in the pre-edge region. Since *(i)* self-absorption corrections were found to be weak in the pre-edge range, *(ii)* all samples were measured in the same conditions and *(iii)* all MCD spectra are normalized to the pre-edge maximum, we assume that self-absorption effects do not impact our analysis.

**Supporting Information**
XRD patterns, soft XAS/XMCD data, RIXS/RIXS-MCD data of reference samples, magnetization curves measured by VSM and soft XMCD.

**Acknowledgements**
The authors thank Hugo Vitoux from the Sample Environment Support Service of the ESRF for his technical support and Jacques Borrel from the mechanical engineering pool of ESRF for his support in the assembly of the experimental setup. The beamlines ID24/BM23 of ESRF are also acknowledged for borrowing their electromagnet. The authors are very grateful to Dr. Edwige Otero from SOLEIL who has implemented on the DEIMOS beamline the 2 Tesla electromagnet (MK2T) end station used for the soft XMCD measurements. Benoit Baptiste is thanked for fruitful discussions on XRD.

This work was supported by French state funds management by the ANR within the *Investissements d'Avenir programm* under reference ANR-11-IDEX-004-02, and more specifically within the framework of the Cluster of Excellence MATISSE. M.S. acknowledges support from the National Science Centre of Poland (2014/14/E/ST3/00026).

A.J. and N.D. initiated and conceived the project. A.J. supervised the project. MV.G. and S.N. implemented the synthetic and analytical experiments. V.G. performed the TEM imaging. N.D performed the powder X-ray diffraction. N.D., V.G. and V.D. performed the magnetomety measurement and N.D. analyzed the magnetometry measurements. A.J., M.S., M.R., Ph.S., F.C., N.B., and N.D. performed the RIXS-MCD measurements. M.S. and A.J. analyzed the RIXS-MCD measurements, M.S. implemented the code for the RIXS-MCD map analysis. N.D, F.C, M.A.A and Ph. O. performed and analyzed the soft-XAS and soft-XMCD measurements. D. T. and A. G. performed the EELS analysis. A.J. and N.D. wrote the manuscripts with the help of all authors. All authors have given approval to the final version of the manuscript.

# Supporting Information

**Nanoscale distribution of magnetic anisotropies in bimagnetic soft core-hard shell MnFe$_2$O$_4$@CoFe$_2$O$_4$ nanoparticles**

*Niéli Daffé, Marcin Sikora, Mauro Rovezzi, Nadejda Bouldi, Véronica Gavrilov, Sophie Neveu, Fadi Choueikani, Philippe Ohresser, Vincent Dupuis, Dario Taverna, Alexandre Gloter, Marie-Anne Arrio, Philippe Sainctavit, and Amélie Juhin*\**

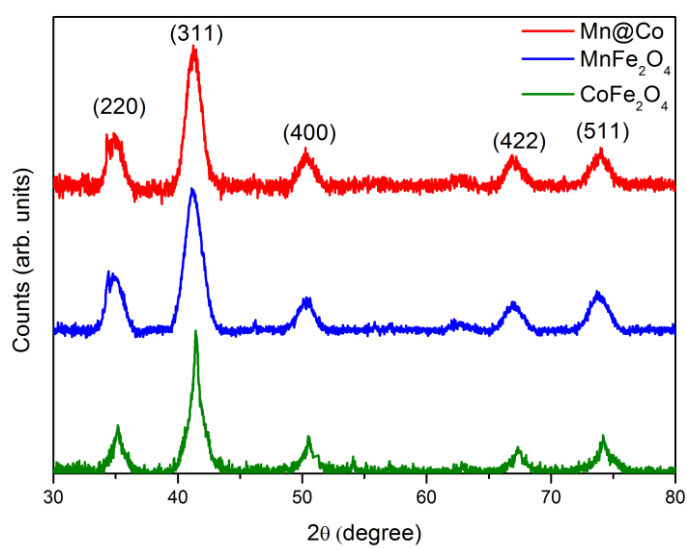

**Figure S1.** XRD pattern of the CoFe$_2$O$_4$ (green line), MnFe$_2$O$_4$ (blue line) and Mn@Co (red line) nanoparticles samples.

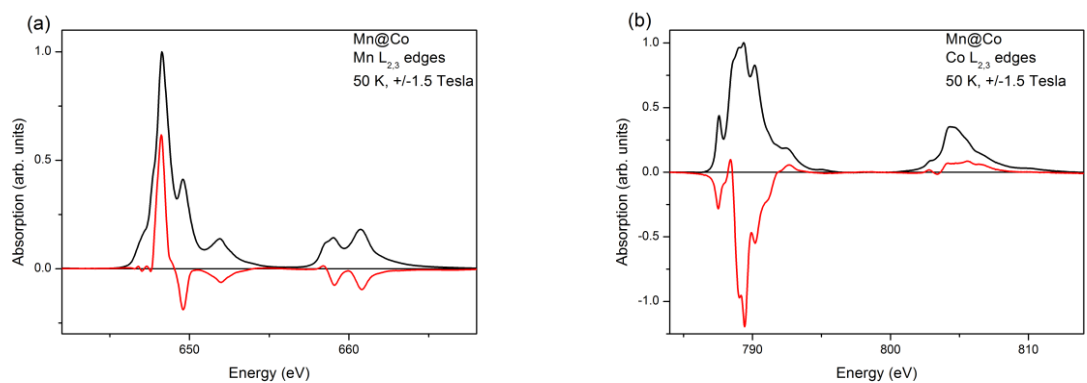

**Figure S2.** Soft XAS and XMCD spectra measured on a dropcast of Mn@Co nanoparticles at T= 50K: (a) Mn L$_{2,3}$ edges, (b) Co L$_{2,3}$ edges.

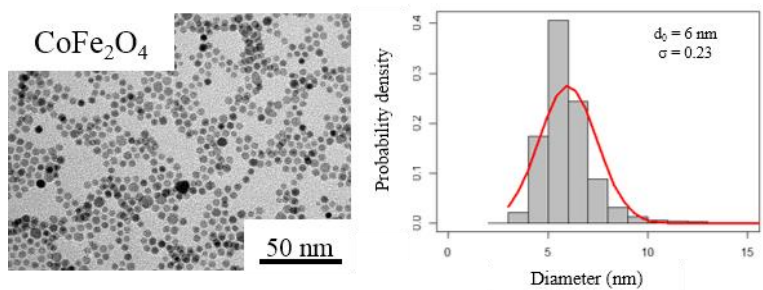

**Figure S3.** TEM image of the CoFe$_2$O$_4$ reference sample and size histogram.

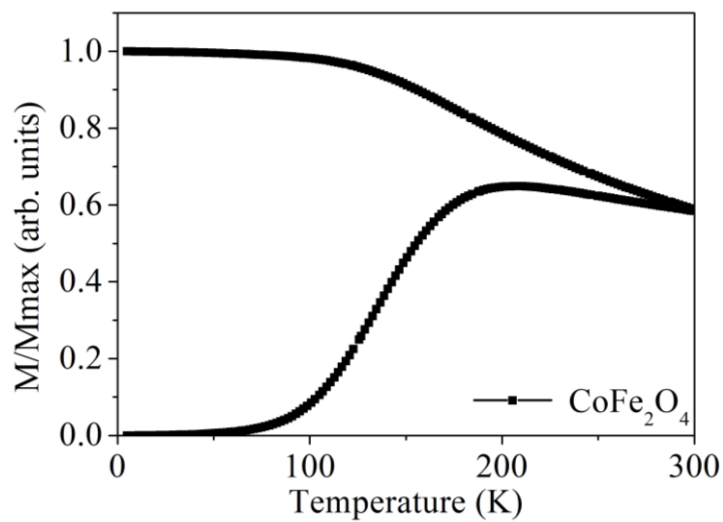

**Figure S4.** Field Cooled and Zero Field Cooled curves measured on CoFe$_2$O$_4$ reference sample (black line).

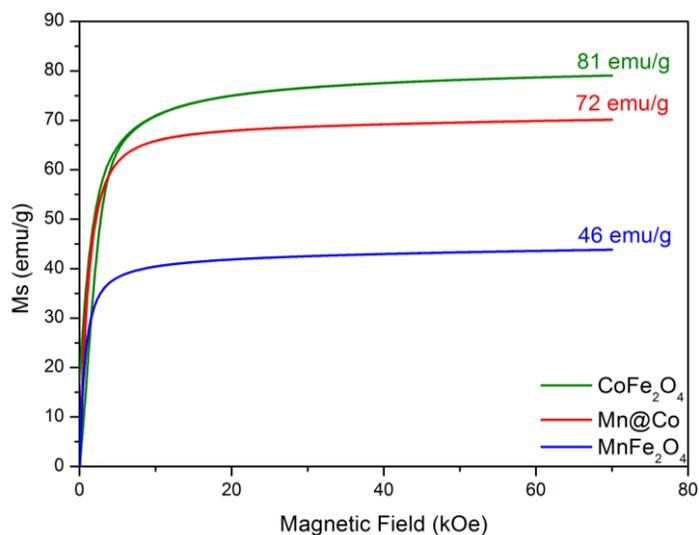

**Figure S5.** VSM magnetometry measurements at 300K of MnFe$_2$O$_4$, CoFe$_2$O$_4$ and Mn@Co powder nanoparticles and saturation magnetization values.

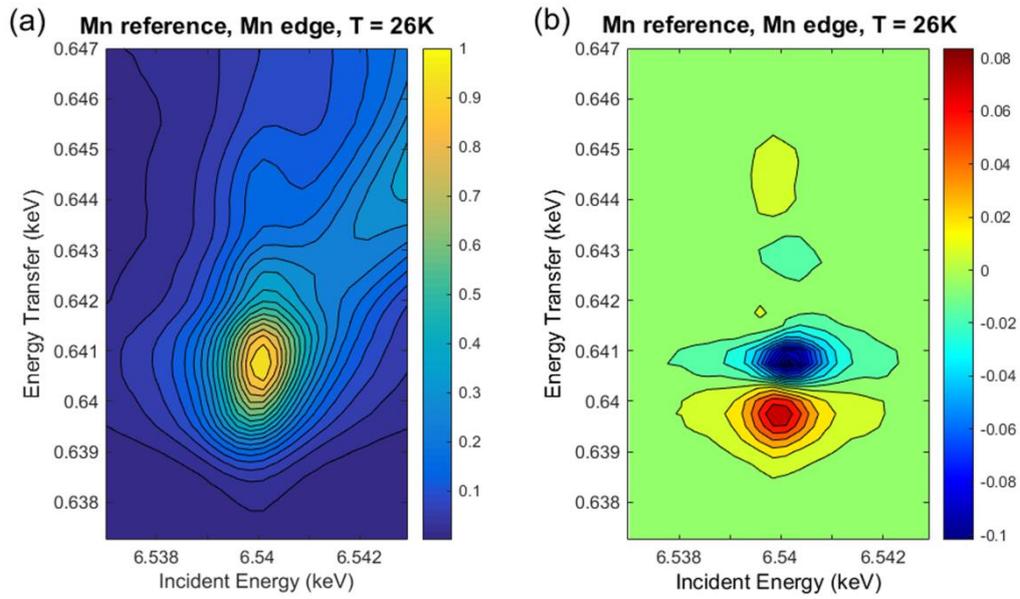

**Figure S6.** RIXS (panel (a)) and RIXS-MCD planes (panel (b)) measured at T=26K on the frozen phase of MnFe$_2$O$_4$ ferrofluid reference.

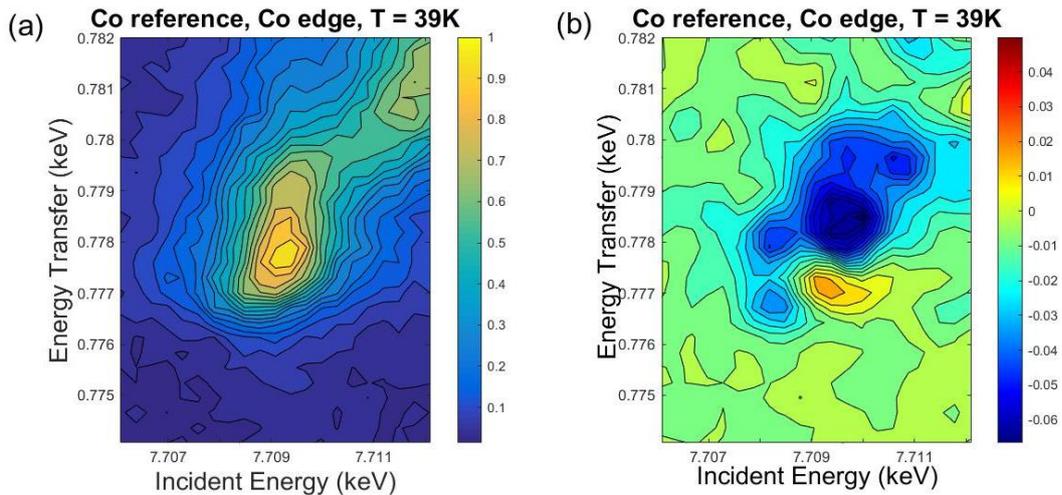

**Figure S7.** RIXS (panel (a)) and RIXS-MCD planes (panel (b)) measured at T=39K on the frozen phase of CoFe$_2$O$_4$ ferrofluid reference.

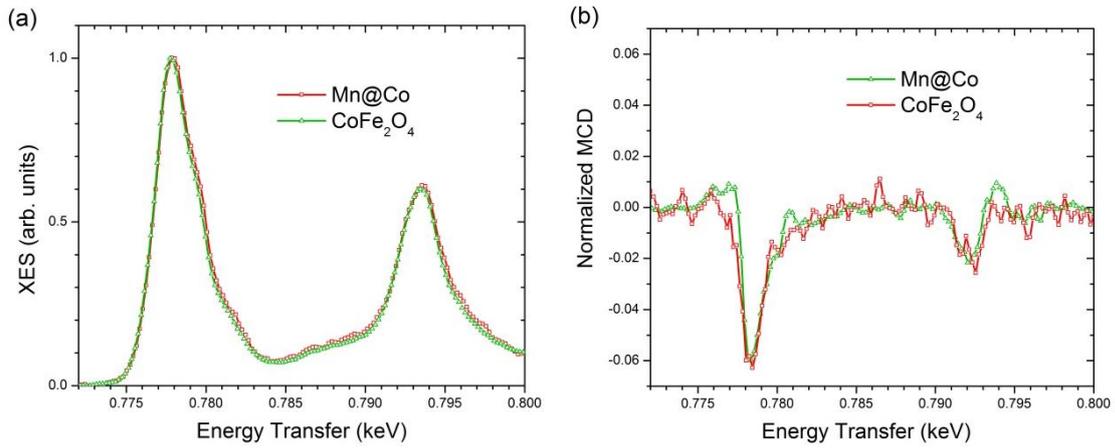

**Figure S8.** Low Temperature XES (panel (a)) and XES-MCD spectra (panel (b)) measured at Constant Incident Energy on the frozen phases of $CoFe_2O_4$ ferrofluid and Mn@Co ferrofluid.

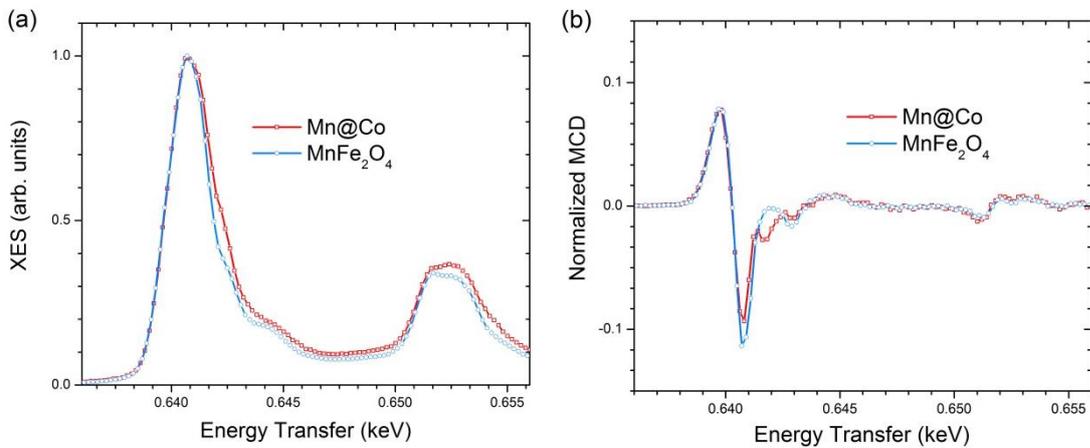

**Figure S9**. Low Temperature XES (panel (a)) and XES-MCD spectra (panel (b)) measured at Constant Incident Energy on the frozen phase of $MnFe_2O_4$ ferrofluid and Mn@Co ferrofluid.

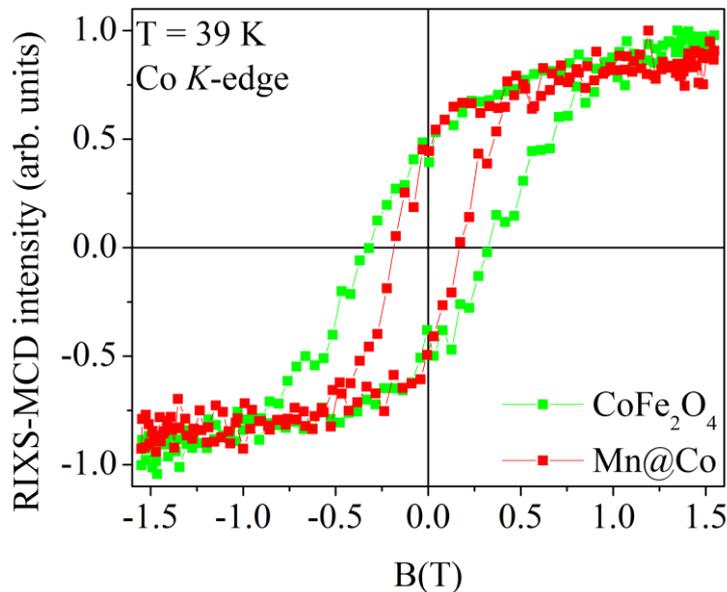

**Figure S10.** Element selective magnetization curves measured at 39K by RIXS-MCD using bulk sensitive hard X-rays at the Co edge for the 6 nm $CoFe_2O_4$ reference sample (green line) and the Mn@Co nanoparticles (red line).

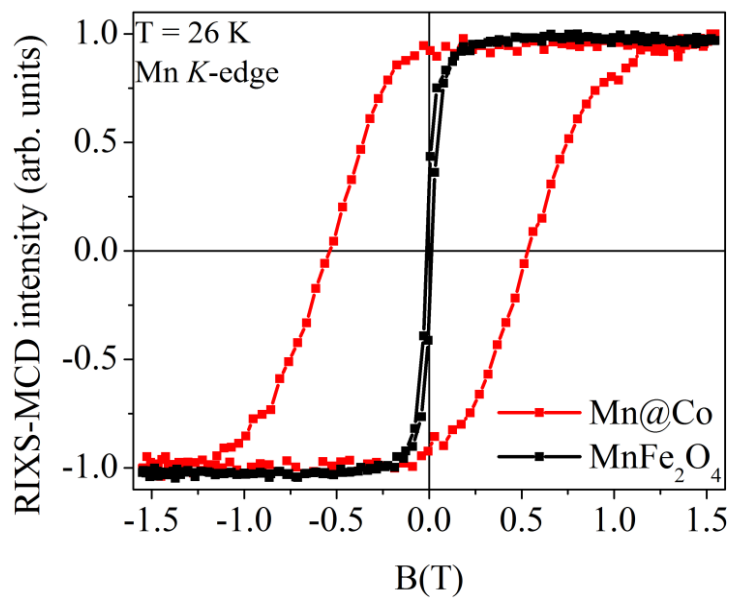

**Figure S11.** Element selective magnetization curves measured at 26K by RIXS-MCD using bulk sensitive hard X-rays at the Mn edge for the MnFe$_2$O$_4$ reference sample (blue line) and the Mn@Co nanoparticles (red line).

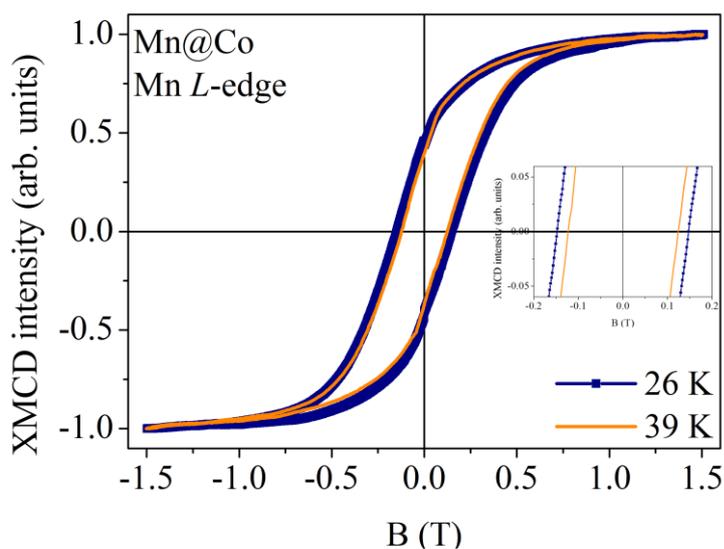

**Figure S12.** Element selective magnetization curves measured by XMCD using surface sensitive soft X-rays at the Mn edge at 26 K (blue line) and at 39 K (orange line).

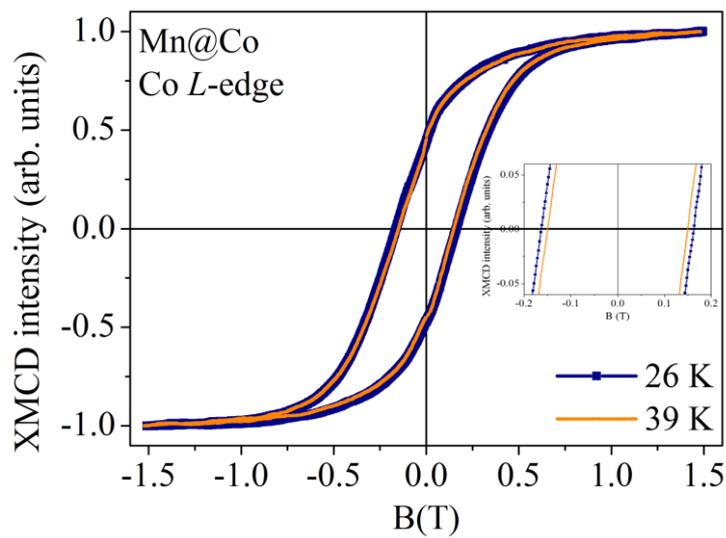

**Figure S13.** Element selective magnetization curves measured by XMCD using soft X-rays at the Co edge at 26 K (blue line) and at 39 K (orange line).